\documentclass[twocolumn,prb]{revtex4}

\usepackage{graphicx}
\usepackage{dcolumn}
\usepackage{bm}

\newcommand{\be}{\begin{equation}}
\newcommand{\ee}{\end{equation}}

\newcommand{\PP}{{\mathcal{P}}}
\newcommand{\HH}{{\mathcal{H}}}

\newcommand{\bsigma}{{\mathbf \sigma}}
\begin{document}

\title{The Magnetic Ordering of the 3d Wigner Crystal}

\author{Ladir C\^andido }
\affiliation{Instituto de F\'isica, Universidade Federal de
Goi\'as, Campus Samambaia, 74001-970, Goi\^ania GO, Brazil}

\author{B. Bernu}

\affiliation{Laboratoire de Physique Th\'eorique des Liquides, UMR
7600 of CNRS, Universit\'e P. et M. Curie, boite 121, 4 Place
Jussieu, 75252 Paris, France}

\author{D. M. Ceperley}
\affiliation{Dept. of Physics and NCSA, University of Illinois at
Urbana-Champaign, Urbana, IL 61801} \email{ceperley@uiuc.edu}
 \vspace*{3mm}

\begin{abstract}
Using Path Integral Monte Carlo, we have calculated exchange
frequencies as electrons undergo ring exchanges  of 2, 3 and 4
electrons in a ``clean'' 3d Wigner crystal (bcc lattice) as a
function of density. We find pair exchange dominates and estimate
the critical temperature for the transition to antiferromagnetic
ordering to be roughly $1 \times 10^{-8}$Ry at melting. In
contrast to the situation in 2d, the 3d Wigner crystal is
different from the solid bcc $^3$He in that the pair exchange
dominates because of the softer interparticle potential. We
discuss implications for the magnetic phase diagram of the
electron gas.
\end{abstract}

\maketitle

\section{Introduction}

The uniform system of electrons is one of the basic models of
condensed matter physics. Wigner\cite{wigner} pointed out that at
low density, the potential energy dominates and the system will
form what is now called a Wigner crystal (3dWC). There have been
attempts to make laboratory examples of the low density
homogeneous 3d electron gas with specially designed
band-engineered AlGaAs heterostructures\cite{super}.
In this paper, we report on calculations of the spin Hamiltonian
in the low density 3d Wigner crystal.

At $T=0$, the properties of the electron gas are determined by a
single dimensionless parameter $r_s = a/a_0 =( m^*/m\epsilon )a/
a_b$ where $a=(3/4\pi\rho)^{1/3}$, $a_b$ is the Bohr radius, $m^*$
is the effective mass, and $\epsilon$ the dielectric constant. In
this paper we will use effective Rydbergs for energies
Ry$^*=(m^*/m_e \epsilon^2)$Ry and $a$ for units of length. In
these units the Hamiltonian is:
 \be\label{fullH}
 H= - \sum_i \frac{1}{r_s^2} \nabla_i^2  + \frac{ 2}{r_s}\sum_{i,j}
 \frac{1}{r_{i,j}}.
 \ee

Though a variety of methods have been applied to calculate the
properties of the low density electron system, the most successful
have been direct simulation methods: namely Quantum Monte Carlo.
Ceperley and Alder\cite{CA80} using Diffusion Monte Carlo (DMC)
determined that melting at zero temperature occurs at $r_s \simeq
100 \pm 20$ for spin 1/2 fermions, and at $r_s \simeq 160$ for
bosons. (This is also the zero temperature melting density for
distinguishable particles). Later estimates\cite{ortiz} found
melting at higher densities ($r_s = 65\pm 10$), but the
variational trial functions in the liquid phase were not
sufficiently accurate. Very recently, Drummond {\it et
al.}\cite{needs04} confirmed the estimate of $r_s = 106 \pm 1$
using DMC with a variety of better functions and the more accurate
results of Zong {\it et al.}\cite{zong02} in the fluid phase. The
bcc crystal structure has the lowest energy throughout the
stability region of the crystal. See, for example, the QMC
calculations of Harris {\it et al.}\cite{ortiz} who found bcc the
lowest energy structure in the range $r_s>60  $.

Once the melting density is established, it is of interest to
determine the low temperature spin order. Harris {\it et
al.}\cite{ortiz} reported finding the ferromagnetic bcc phase as
stable, however, other aspects of those calculations have not been
reproduced. Drummond {\it et al.}\cite{needs04} attempted to
determine directly the energy difference between ferromagnetic and
antiferromagnetic orderings using DMC but found the difference
zero within their error bars (on the order of $2 \times 10^{-7}$Ry
at $r_s=100$). This is consistent with the results reported below.
One needs to use a method sensitive to the small magnetic
energies, which are typically many orders of magnitude smaller
than the plasmon energies that determine the accuracy of the DMC
energies.

\begin{figure}
\includegraphics[width=8cm,height=8cm]{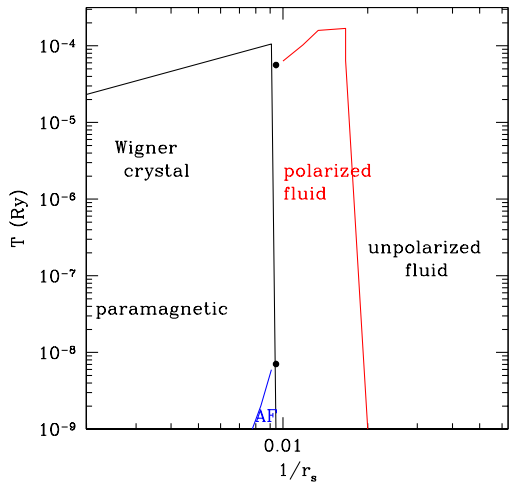}
 \caption{\label{pdrs}
Phase diagram of the 3d electron gas showing the region of
stability of the crystal\cite{jones}, the polarization transition
from QMC calculations\cite{zong02} and the antiferromagnetic
transition (this work). (color online)}
\end{figure}

Jones and Ceperley\cite{jones} studied the quantum melting curve
for distinguishable particles. At densities for $r_s \geq 100$ the
melting is classical, and occurs for temperatures\cite{cocp} $k_B
T_{melt} = 2 /(\Gamma_c r_s)$Ry where $\Gamma_c \approx 173$. We
only use this melting to determine the region where the crystal is
stable, and thus the region where magnetic ordering is relevant.
Fig. \ref{pdrs} summarizes the 3deg phase diagram.

For spin 1/2 particles, the magnetic ordering is not fixed by the
spatial ordering. The earliest quantitative calculation was using
a Slater determinant of localized Wannier functions ({\it i. e.}
Gaussians) by Carr.\cite{carr} He found an antiferromagnetic phase
in the Wigner crystal at intermediate density and ferromagnetic at
lower density. We will compare with this calculation later in the
paper. Edwards and Hillel\cite{edwards} using a Hartree-Fock
method with a flexible, delocalized basis and a variety of spin
orderings found antiferromagnetic ordering at intermediate density
$10<r_s<40$ and ferromagnetic at lower densities. But, as
mentioned above, the crystal phase is only stable for $r_s > 106$.

Herring\cite{herring} reviewed the situation as understood in 1965
in some detail, including the contribution of ring exchanges of
electrons.  Thouless\cite{thouless} introduced the current theory
of magnetism in quantum crystals. According to this theory, in the
absence of point defects, at low temperatures the electrons will
almost always be near a lattice site. If the system is constrained
to stay in the neighborhood of the two perfect lattice positions
$Z$ and $PZ$ where $P$ is a permutation of particle labels, the
exchange frequency equals the splitting between the antisymmetric
spatial state and the symmetric spatial state: $2J_P= E_A-E_S >0$.
The spin Hamiltonian comes about from making the total
wavefunction antisymmetric:
 \be
\HH_{spin} = - \sum_P (-1)^P J_P {\hat \PP}_{spin}\label{spinh}
 \ee
where the sum is over all cyclic (ring) exchanges described by a
cyclic permutation $P$, and ${\hat \PP}_{spin}$ is the
corresponding spin exchange operator. (Although more complex
products of several ring exchanges are possible, in cases
considered, they are negligible.) The sign, $(-1)^P$, implies that
an exchange of even number of electrons is antiferromagnetic and
an odd number of electrons is ferromagnetic.  Ring exchange models
have been used to describe correlated electron systems, such as
high temperature superconductors\cite{hightc}, quantum Hall
systems\cite{kivelson} as well as electrons\cite{2dWC} and helium
atoms\cite{rogerhe} confined in planes.

One might expect that pair electron exchanges would dominate over
higher-body exchanges. Since the bcc lattice is  bipartite, a
simple N\'{e}el antiferromagnetic state would seem to be favored.
Rather surprisingly, it has been found\cite{RHD} that in  3d solid
$^3$He, which also forms a bcc lattice, exchanges of 2, 3 and 4
particles have roughly the same order of magnitude and must all be
taken into account to understand the magnetic ordering. This is
known as the multiple spin exchange model(MSE). The resulting spin
order is more complex since the order is frustrated by the
competing exchanges. We wish to determine whether such a model is
relevant for the 3dWC.

\begin{figure}
\includegraphics[width=8cm,height=8cm]{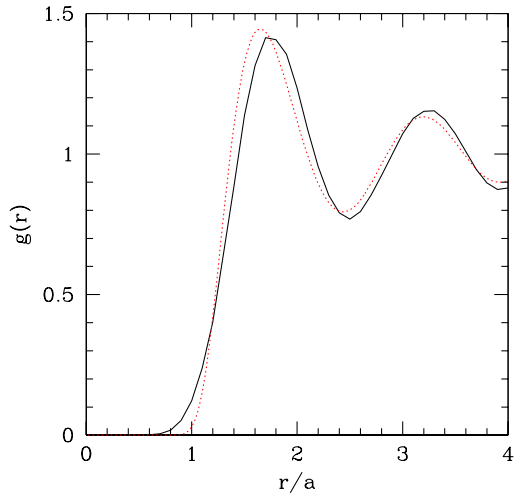}
 \caption{\label{gofr}
The pair correlation function for the 3dWC (solid line)
($r_s=100$) and for bcc $^3$He (dashed red line)also at melting
($24.23cm^3/mole$. The functions are nearly identical, though the
electrons can get significantly closer together than helium atoms
can. (color online) }
\end{figure}

Figure \ref{gofr} shows the pair correlation functions for solid
$^3$He and the electron gas near the crystallization density.
Because the $g(r)$'s are so similar, one might expect their
exchanges frequencies would be similar and hence have the same
magnetic ordering. We also note that the Lindemann's ratio, the
mean squared displacement in units of the nearest neighbor
spacing, for bulk helium and the Wigner crystal are also similar
near melting ($0.32$ and $0.30$ respectively). However, note the
$g(r)$'s are very different at small $r$ because the potentials
are so different; the helium-helium interaction is much more
repulsive at short distances.

In this paper, we determine the magnetic interaction in the Wigner
crystal, based on Thouless'\cite{thouless} theory of exchange.
Path Integral Monte Carlo (PIMC) as suggested by
Thouless\cite{thouless} and Roger\cite{roger84} has proved to be a
reliable way to calculate these parameters directly from the
Coulomb interaction. The theory and computational method have been
tested thoroughly on the magnetic properties of bulk helium
obtaining agreement with measured properties\cite{RMPI}.
We\cite{2dWC} have also used this method to calculate exchange
frequencies in the 2dWC. Here we report results for the 3dWC.

\section{Computational Details}

The Path Integral method for calculating exchange frequencies is
based on the ratio:
 \be
f_P (\beta) = \frac { \langle Z | e^{-\beta H}  | P Z \rangle}
 { \langle Z | e^{-\beta H}  | Z \rangle}\label{fdef}
  \ee
where $Z$ represents the many-body configuration of electrons
sitting on the bcc lattice sites and $PZ$ is a permutation of
those sites. Then under general assumptions, the exchange
frequencies are given by:
 \be
 f_P (\beta) = \tanh ( J_P (\beta - \beta_0) )\label{tanh}
 \ee
where  $\beta_0$ is the amount of imaginary time to initiate the
exchange. The ratio $f_P$ is determined by a specialized Path
Integral Monte Carlo method and Eq. \ref{tanh} is inverted to
determine $J_P$.

We do simulations with two types of paths: a) paths beginning and
ending at the perfect lattice positions and b) paths beginning at
$Z$ and ending at $PZ$. The imaginary time density matrices
$e^{-\beta H}$ are expanded out into a path integral connecting
the end-points of the paths. Using the polymer
``isomorphism''\cite{RMPI}, $f_P(z)$ is related to the free energy
need to induce a specific cross-linking $P$ into a crystal of ring
polymers.  Using the Bennett method\cite{bennett,CJ}, we directly
determine the ratio $f_P$ by examining histograms of the change in
action in mapping paths of one type into paths of the other type.
With this method we can determine very small frequencies to an
accuracy of a few percent, irrespective of the magnitude of $J_P$.

Ewald sums are used to represent the Coulomb interaction in
periodic boundary conditions, taken as a cube.  The potential is
split into long-range and short-range terms with the usual
Gaussian breakup\cite{ceperley78}. Then the exact two particle
action for the short ranged potential (the complementary error
function) is determined using matrix squaring.\cite{RMPI} The long
range action is taken in the primitive approximation; this is
appropriate since it is smooth. Most calculations were done with
54 electrons in the simulation cube with a few tests of 128
electrons giving agreement. Note that at low density, long
wave-length plasmons are strongly suppressed by the potential
energy, making the exchange more localized spatially than is the
case with solid helium.

One numerical approximation is the imaginary-time step, or
equivalently, the number of points on the path. We did several
calculations for each value of $r_s$ and each type of exchange to
establish that the results are in the zero time step limit within
error bars.  A second approximation concerns the value of $\beta$
in Eq. \ref{fdef}. Because the exchanges are instantons ({\it i.
e.} confined in imaginary time), the results converge quickly in
$\beta$  as long as $\beta_0 < \beta$.  We typically choose $\beta
< 3\beta_0$ and observe very weak dependance on $\beta$.
Typically, for converged results, this implies one to two hundred
steps in the exchange. Further details of the method have been
discussed in earlier papers.\cite{cornell}

In order to keep the system from melting for densities higher than
the bosonic melting\cite{CA80} $100 \leq r_s \leq 160$, the
electron paths are restricted to stay in the positive region of a
trial wavefunction: the fixed-node boundary conditions. We
separate the electrons into those exchanging (say ``$p$'' of them)
and the spectators ( {\it i. e.} $N-p$ electrons). A Slater
determinant of only the spectator electrons is constructed:
 \be
\psi (R(t)) = \det [ \exp(-c ({\bf r}_i (t) - {\bf z}_j)^2 ) ]
 \ee
where ${\bf z}_j$ is the set of spectator lattice sites and ${\bf
r}_i(t)$ the imaginary time path of the spectator electrons. The
value of the parameter $c$ was optimized by variational Monte
Carlo\cite{ceperley78}: $c = 0.2 r_s^{1/2}$. The more recent
values obtained by minimizing the fixed-node energy\cite{needs04}
$c = 0.11 r_s^{1/2}$ were not available when we did the
calculations. A Slater determinant of Gaussian orbitals is an
accurate representation for the nodes of the many-electron ground
state wavefunction. More complicated forms, such as linear
combinations of Gaussians, do not result in a significantly better
trial functions\cite{needs04} for the 3dWC.

Only paths which keep a positive determinant throughout the path
are kept $\psi(t) > 0$ for $ 0\leq t\leq \beta$: this is the
fixed-node method. We find that these boundary conditions are
sufficient to keep the system from melting. We do see a
suppression of the exchange frequency caused by the determinantal
boundary conditions at $r_s=150$ of about $10\%$. In principle,
the spin ordering should be determined self consistently. For
example, it would be better to apply antiferromagnetic boundary
conditions, however, we have not tested that approach. For this
reason, we may have corrections to the exchange frequencies on the
order of 10\%, particularly near melting.

\section{Exchange Frequencies}

We have calculated 2, 3 and 4 particle exchanges for six different
densities in the range $100 \leq r_s \leq 150$.  We determined two
different pair exchanges: first neighbor, $J_{nn}$, and second
neighbor, $J_{nnn}$. We do find a significant 2 body
next-nearest-neighbor exchange. The most compact three-body
exchange, $J_t$, has two first-neighbor and one second-neighbor
electrons. There are two types of four electron exchanges
involving only first neighbors\cite{RHD}: the planar exchange,
$J_p$, and the folded exchange, $J_f$. Table I shows the exchange
frequencies of our calculations versus density.

\begin{figure}
\includegraphics[width=8cm,height=8cm]{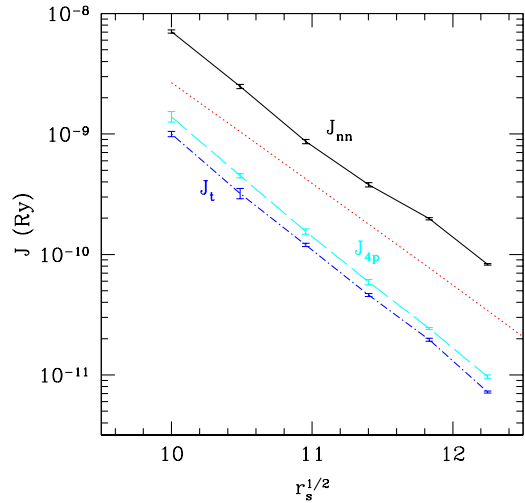}
 \caption{\label{J2}
(color online) Exchange frequencies (in Ry log scale) versus
$r_s^{1/2}$. The solid (black) line shows the 2-body nearest
neighbor exchange. The dashed (red) line is from the calculations
of Carr\cite{carr}. The other lines are the triple (blue short
dashes) and planar 4-electron (cyan, long dashes) exchange
frequencies. A WKB calculation would be a straight line. }
\end{figure}

Figure \ref{J2} shows the exchange frequencies versus density. We
find that the exchange energies are very much smaller than the
plasmon energies. The kinetic energy of the Wigner crystal can be
expanded in a power series\cite{ceperley78}:
 \be
  T= -\frac{d(r_s E)}{dr_s} = 0.669 r_s^{-3/2} -0.553 r_s^{-2}.
 \ee
In the density range of consideration, the kinetic energy varies
from 0.6 mRy to 0.3 mRy: five orders of magnitude greater than the
exchange frequencies.  Roughly speaking, the electrons vibrate
around their lattice sites $10^5$ times before exchanging with a
nearby electron. This is comparable with the situation in bcc
$^3$He and justifies that we can reduce the original Hamiltonian
involving charges, Eq. \ref{fullH},  to the spin Hamiltonian, Eq.
\ref{spinh}.

We see in Fig. \ref{J2} that the exchange frequencies drop off
exponentially: roughly as $\exp(-S r_s^{1/2})$.  This follows from
assuming a single most probable path for the ring
exchange\cite{roger84} is independent of density. (Note that we do
not make this assumption in the PIMC simulation). It is likely
that WKB calculation of $S$ will give reasonable estimates of the
exchange frequencies as they do for the
2dWC\cite{roger84,hirashima}.

\begin{figure}
\includegraphics[width=8cm,height=8cm]{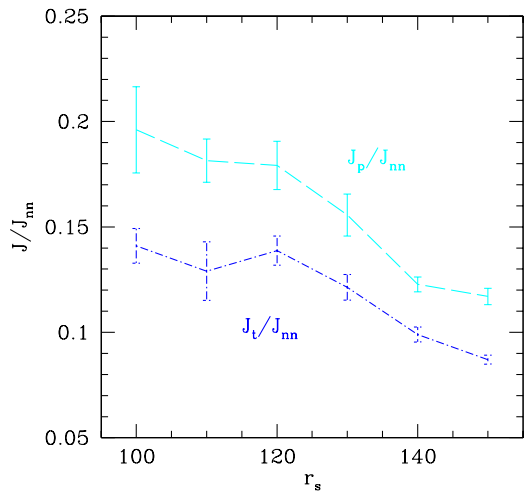}
 \caption{\label{ratio}
(color online) Ratio of exchange frequencies versus $r_s$.  Shown
are $J_t/J_{nn}$ (blue, short dashes) and $J_p/J_{nn}$ (cyan, long
dashes).}
\end{figure}

We find that the rate for pair exchanges is much larger than the
other exchanges, making a very stable antiferromagnetic ground
state. The planar 4-body exchange, also antiferromagnetic, is
slightly larger than the ferromagnetic, 3-body exchange. Only
ratios of the exchange rates can enter into determining the
stability of a given magnetic state. Figure \ref{ratio} shows the
density dependance of the ratios $J_f/J_n$ and $J_t/J_n$. We see a
decrease in these ratios as density decreases, thereby further
stabilizing the antiferromagnetic phase.  Next in importance is
the 2-body next nearest neighbor exchange, followed by the folded
4-body exchange.

\begin{table*}
\caption{Results of the PIMC calculations at various values of
$r_s$. Units are $10^{-9}$ Ry/electron. Quantities in () are the
estimated relative statistical error. The second column is the
estimated transition temperature (also in nRy) based on Eq.
\ref{tc}. The spin couplings $j_n$ are defined in Eq. \ref{effj}.}
\begin{tabular}{|c|c||c|c|c|c|c||c|c|c|}
\hline
  $r_s$&$k_B T_c$&$ J_{nn} $ & $J_{nnn}$& $J_t $ & $J_{p}$ & $J_{f}$ &$j_1$&$j_2$&$j_3$\\

    && 2 (11) &2 (22)& 3 (112)&$4(1^4;23)$&$4(1^4;22)$&&&\\
  \hline
 100  & 16.&7.09 (0.03)  && 1.0 (0.05)  &  1.39 (0.10) & & 5.6&-1.9&0.7  \\
 110  & 6.0&2.48 (0.05)  && 0.32 (0.10)  &  0.45 (0.04)& & 2.0&-0.6&0.2 \\
 120  & 2.0&0.865 (0.04)  &0.058 (0.03) &  0.12 (0.03) & 0.155 (0.04) & 0.0112( 0.05)&0.64&-0.25&0.08 \\
 130  & 0.90&0.379 (0.04) && 0.0460 (0.03) &  0.059 (0.05) &  0.0037 (0.06)&0.29&-0.09&0.03 \\
 140  & 0.49&0.200 (0.04) && 0.0196 (0.03) &  0.0243 (0.02)& &  0.16&-0.04&0.01 \\
 150  & 0.21&0.0827 (.014) && 0.00720 (0.02) &  0.00968 (0.03) &  0.000565 (0.025)&0.071&-0.014&0.005 \\
\hline
\end{tabular}
\end{table*}

\section{Magnetic properties}

Having determined the exchange frequencies, we can now discuss the
magnetic properties, based on the spin Hamiltonian in Eq.
\ref{spinh}. In principle, one has a formidable many-body problem.
However, we can make use of the extensive results available from
studies of solid $^3$He, also a bcc lattice. In particular, see
the review of Roger {\it et al.} \cite{RHD} (RHD) and references
therein. For spin 1/2 particles, one can write the spin
Hamiltonian in terms of Pauli spin matrices. Two and 3 particle
exchanges map into a Heisenberg model:
 \be
   H_h =\frac{j_1}{2}\sum_{i,j}^{(1)} \bsigma_i \cdot \bsigma_j
   +\frac{j_2}{2}\sum_{i,j}^{(2)} \bsigma_i\cdot \bsigma_j + \ldots \label{espinh}
 \ee
where the sums are over first, second and third neighbor pairs
respectively. (Note on notation: $J_x>0$ refers to a ring exchange
frequency for the cycle $x$, while $j_n$ refers to the coupling
constant in Eq. \ref{espinh} between two spins a distance $x$
apart; all frequencies have an opposite sign from the notation of
RHD.) These spin couplings are given in terms of the ring
exchanges by:
 \begin{eqnarray}\label{effj}
    j_1 &=& J_{nn}+3(-2J_t +J_p+J_f) \\\nonumber
    j_2 &=& 2(-2J_t+J_f)+J_p + J_{nnn}\\\nonumber
    j_3 &=& J_p/2.
 \end{eqnarray}
The pair couplings, shown in Table I, can either be positive
(antiferromagnetic) or negative (ferromagnetic) depending on the
relative importance of even and odd ring exchanges.

The four spin exchanges lead to additional four spin terms in the
effective Hamiltonian in Eq. \ref{espinh}.
 \be
    G_{ijkl}= (\bsigma_i\cdot\bsigma_j)(\bsigma_k\cdot\bsigma_l)
    +(\bsigma_i\cdot\bsigma_l)(\bsigma_j\cdot\bsigma_k)
    -(\bsigma_i\cdot\bsigma_k)(\bsigma_j\cdot\bsigma_l)
 \ee
 \be
    \Delta H_x = \frac{J_x}{4}\sum_{i,j,k,l}  G_{ijkl}.
 \ee
The summation ($x=p$ or $x=f$) is over distinct labels describing
the planar or folded four-particle exchanges. According to RHD,
one can neglect these additional terms in estimating properties at
high temperatures. Since they are fourth order in the order
parameter field, they can only contribute at lower temperature.

Given the above spin Hamiltonian, several things can be easily
computed. At high temperature, the Curie-Weiss constant measures
the leading term in a $1/T$ expansion to the susceptibility
($\chi^{-1} = C(T-\Theta + B/T + \ldots )$. We find that:
 \be
   \Theta = 4 J_{nn} + 3 J_{nnn}-36 J_t +18 K_f + 18 K_p.
 \ee
In regions above the ordering temperature, this determines the
deviation with respect to uncoupled spins.  It is positive
(antiferromagnetic), for all densities.

From the dominance of $J_{nnn}$ over the other ring exchanges, we
expect ordering into an antiferromagnetic phase via a second order
transition at a sufficiently low temperature. The complications
resulting from  a frustrated spin Hamiltonian are absent, so that
we can be considerably more confident in our predictions than is
the case with a frustrated spin-model such as helium or the 2dWC.
First, let us determine the transition temperature within mean
field theory. RHD\cite{RHD} (Fig. 16) show the mean field phase
diagram as a function of $(j_1,j_2,j_3)$ As mentioned above, the
four spin terms will not change this diagram. Based on the values
in Table I, we find that the 3dWC is very much in the
antiferromagnetic region. The density dependance shown in Fig.
\ref{ratio}, has a tendency to enhance even more the
antiferromagnetism at larger values of $r_s$.

Within mean field theory, the antiferromagnetic phase has a
critical transition temperature:
 \be
  k_B T_c = 4j_1 -3 j_2 -6 j_3.
 \ee
We can include fluctuations into this estimate by using results of
series expansions. Oitmaa and Zheng\cite{oitmaa} have calculated
properties of the $j_1,j_2$ model on a bcc lattice using Pad\'{e}
approximates to high temperature expansions (up to tenth order in
the inverse temperature.) They find the antiferromagnetic phase is
stable at low temperatures for $j_2 \leq 0.705 j_1$. They also
find that the transition temperature is $k_B T_c \approx 2.76 j_1
- 2.61 j_2$. The effect of fluctuations is to lower $T_c$ by about
30\%. Since the effect of $j_2$ and $j_3$ is to couple the same
sublattices, a reasonable way to extend these results to $j_3\neq
0$ is to assume that the effect of $j_3$ is determined by the
number of spins coupled. Since there are twice as many third
neighbors as second neighbors, we assume:
 \be
  k_B T_c \simeq 2.76 j_1 - 2.61 (j_2 + 2 j_3).
  \label{tc} \ee
The lower line on Fig. \ref{pdrs} and the second column of Table I
show this estimated transition temperature.

For $r_s = 120$, where we have done the most calculations, we find
an estimated transition temperature of $T_c=2.0$ nRy. We can
parameterize the density dependance of the transition temperature
by assuming the WKB form for the dependance of the exchanges on
the density. We obtain $k_B T_c \approx 3.7 \exp( -1.92
r_s^{1/2})$ Ry. (Note: because we  calculated $J_{nnn}$ and $J_f$
only at $r_s=120$, we estimated their values at other densities by
scaling. The importance of these exchanges is negligible.)

We now compare the present calculation with DMC calculations aimed
at determining the spin ordering. This is done by performing a
ground state total energy calculation (within the fixed node
method) for a fully polarized Wigner crystal and for an
antiferromagnetic crystal. Such an attempt was made both for the
2dWC\cite{zhu} and for the 3dWC\cite{needs04} but without finding
a significant energy difference. Given the exchange frequencies we
can estimate this energy difference. Assuming classical spins, the
antiferromagnetic and ferromagnetic energies are easily found to
be:
 \begin{eqnarray}
    E_{AF}/N  &=& -2 j_1 + 1.5 j_2 + 3 j_3\\
     E_F/N &=& 2 j_1 + 1.5 j_2 + 3 j_3.
 \end{eqnarray}

Oitmaa and Zheng\cite{oitmaa} determine corrections to the
classical antiferromagnetic ground state energy of the $(j_1,j_2)$
model using an expansion technique and find:
 \be
  E_{AF}/N = -2.3 j_1 + 1.32  j_2.
 \ee

This is only roughly 10\% lower than the classical value for the
3dWC parameters. Note that the classical ferromagnetic energy
needs no correction. Using these estimates and Eq. \ref{effj},
within a few percent, the energy difference to spin polarize the
crystal is:
 \be
    \Delta E \approx (E_F - E_{AF})/N =4.3 j_1 +0.18(j_2+2j_3).
 \ee
Near the melting density of $r_s = 100$, $\Delta E \approx 2.4
\times 10^{-8}$ Ry which is a factor of 10 smaller than the
reported\cite{needs04} statistical errors within DMC. Even if one
could reduce the statistical error, the usual spin wavefunction in
the antiferromagnetic phase is relatively crude, and the nodal
surfaces are not necessarily optimized for the regions where
exchanges occurs. Calculation of the ring exchange frequencies are
a much more direct and efficient way of determining the magnetic
ordering, and yield more physical insight into the microscopic
mechanisms giving rise to the magnetism.


Concerning previous calculations on the magnetic properties of the
bcc Wigner crystal at low temperature, Carr\cite{carr} did a
calculation of the 3d Wigner crystal and determined the harmonic
energy and the two electron exchange integral. He did this
assuming the ground state wave function is a product of single
Gaussian orbitals and estimated the exchange frequency by
calculating the exchange matrix element obtaining:
 \be
   J_{nn} \approx  ( 1.6 r_s^{-.75} -6.5 r_s^{-1} ) \exp(-1.55
   r_s^{1/2} ).
 \ee
As shown in Fig. \ref{J2}, the slope of the exchange coefficient
is roughly correct, while the prefactor is too small by a factor
of about three. By including all effects of electron correlation,
the tunnelling frequency is enhanced over what one gets with an
uncorrelated wavefunction. More seriously, because of the two
terms of opposite sign, he found an antiferromagnetic to
ferromagnetic transition at $r_s =270$. Carr's work is previous to
the tunnelling theory of Thouless\cite{thouless} and
Herring\cite{herring}, which asserts that purely pair exchanges
must be antiferromagnetic, so a sign change like this can only
come about from cancellation of exchanges of even and odd number
of electrons. In Carr's calculation, there is no consideration of
the possibility of more than two electron exchanging. The sign
change comes from neglecting correlation between the electrons
that are exchanging.\cite{herring}  Given the strong correlations
present at $r_s \geq 100$, it is remarkable that the order of
magnitude is reasonable.  A reasonable estimate would not be
obtained this way for the solid $^3$He because the hard core
interaction makes the interaction matrix elements infinite, or
nearly so. Carr also calculated the exchange for second neighbors;
at $r_s=120$ he obtained $J_{nnn}/J _{nn} = 0.037$, while we
obtained a ratio of $0.067$.

Now turning to comparison with other quantum systems, similar
behavior\cite{2dWC} in the spin Hamiltonian was found near melting
between the second layer of $^3$He absorbed on graphite and the
2dWC, but the similar behavior does not seem to occur in 3d. It is
suggested that the similarity is due to a common vacancy
interstitial fluctuation mechanism, something that is also related
to the melting of the quantum crystal\cite{C174}, which is
possibly second order or nearly so.

This similarity is not found in 3d. The magnetic properties of the
3dWC are quite different than bcc solid $^3$He. This comes about
because pair exchanges are relatively more important in the Wigner
crystal: we find in the 3dWC that $J_t/J_{nn}=0.14$ while in
$^3$He the ratio is $0.41$. This crucial enhancement of the pair
exchange comes about because of the difference in the interaction
strength at short distances: the helium-helium interaction has a
much harder core. During a pair exchange, the particles have to
pass by each other. As can be seen in Fig. \ref{gofr} electrons
can approach closer than helium atoms, and this allows them to
pass by each other much more frequently. Exchange is a tunnelling
process and the rate is very sensitive to the phase space in the
transverse direction that the electrons have when they undergo
exchange.

The frustration between 2, 3 and 4-body exchanges leads, in solid
helium, to a more complicated N\'{e}el ordered ground state, the
uudd phase\cite{RHD}. However, in the 3d Wigner crystal, pair
exchange dominates leading to a very stable antiferromagnetic
phase. We find that the relative frequency of 3- and 4-body
exchanges is about the same in the two systems: $J_t/J_p= 1.42$ in
$^3$He and 1.27 in the 3dWC. In 3- and 4-body exchanges, the
particles do not approach each other so closely, and the
similarity in $g(r)$ leads to similar ratios of exchange
frequencies.

Recently, it was found in DMC calculations\cite{zong02} that the
3d electron fluid has a partially spin polarized ground state at
low density. In fact, it undergoes a second order transition to a
partially polarized state at $r_s = 50 \pm 2$. The polarization
increases until it is fully polarized at freezing $r_s =106$.
(Note, however, that because of the fermion sign problem, other
types of ordering, such as superfluidity, cannot be ruled out. In
this respect, the liquid is much more difficult to treat than the
crystal, since in the crystal, fermion effects are very small.)
The region of stability of the spin polarized fermi liquid is
shown in Fig. \ref{pdrs}. It is curious that both the fluid and
crystal have magnetic ordering near the melting line. However, in
the crystal, the magnetic ordering occurs at a temperature more
than one thousand times lower than in the fluid. As in the 2dWC,
in the 3dWC the magnetic ordering temperature drops off very fast
as density decreases.

We hope to have provided a definitive result concerning the
magnetic ordering of the 3dWC, a system which has provoked much
speculation over the years. Because of the small energy scales, it
will be an experimental challenge to observe the magnetic ordering
in the 3d Wigner crystal.

\begin{acknowledgments}
This research was supported by NSF DMR01-04399 and the Department
of Physics at the University of Illinois Urbana-Champaign and the
CNRS-University of Illinois exchange program. Computational
resources were provided by the NCSA. LC thanks the support by
Conselho Nacional de Desemvolvimento Cient\'ifico e Tecnol\'ogico
(CNPq).

\end{acknowledgments}

\end{document}